\documentclass[reqno,12pt]{article}
\pdfoutput=1
\usepackage{amsmath,amsfonts,amssymb,amsthm,amstext,amscd}
\usepackage[all]{xy}
\usepackage{hyperref}
\usepackage{epsfig}
\usepackage{graphicx}
\usepackage{tikz}
\usetikzlibrary{decorations.pathmorphing}
\usepackage{caption}
\usepackage{subcaption}
\usepackage{lipsum}
\usepackage{psfrag}
\usepackage{epstopdf}
\usepackage{comment}
\usepackage{cite}

\newcommand{\be}{\begin{equation}}
\newcommand{\ee}{\end{equation}}
\newcommand{\bea}{\begin{eqnarray}}
\newcommand{\eea}{\end{eqnarray}}
\newcommand{\bse}{\begin{subequations}}
\newcommand{\ese}{\end{subequations}}
\newcommand{\beqa}{\begin{eqnarray}}
\newcommand{\eeqa}{\end{eqnarray}}
\newcommand{\beqar}{\begin{eqnarray*}}
\newcommand{\eeqar}{\end{eqnarray*}}
\newcommand{\bi}{\begin{itemize}}
\newcommand{\ei}{\end{itemize}}
\newcommand{\bn}{\begin{enumerate}}
\newcommand{\en}{\end{enumerate}}

\newcommand{\ba}{\begin{array}}
\newcommand{\ea}{\end{array}}
\newcommand{\bc}{\begin{center}}
\newcommand{\ec}{\end{center}}

\newcommand{\phii}{\varphi}

\newcommand{\ie}{{\em i.e.}\ }
\newcommand{\eg}{{\em e.g.}\ }

\newcommand{\p}{\partial}

\newcommand{\de}{\delta}

\newcommand{\vm}{{\vec{m}}}

\newcommand{\vn}{{\vec{n}}}
\newcommand{\vk}{{\vec{k}}}

\newcommand{\NHEG}{{\text{Diff}_\vk(T^{n+1})}}

\newcommand{\eps}{\epsilon}
\newcommand{\bfeps}{\boldsymbol{\epsilon}}

\newcommand{\bL}{{\boldsymbol L}}

\newcommand{\bJ}{\boldsymbol{J}}

\newcommand{\cF}{\mathcal F}
\newcommand{\cG}{\mathcal G}

\newcommand{\cO}{\mathcal O}

\makeatletter \@addtoreset{equation}{section}

\makeatletter\renewcommand\section{\@startsection {section}{1}{\z@}%
                                   {-3.5ex \@plus -1ex \@minus -.2ex}
                                   {2.3ex \@plus.2ex}%
                                   {\normalfont\large\bfseries}}
\renewcommand\subsection{\@startsection{subsection}{2}{\z@}%
                                     {-3.25ex\@plus -1ex \@minus -.2ex}%
                                     {1.5ex \@plus .2ex}%
                                     {\normalfont\bfseries}}

\usepackage[mathscr]{eucal}
\usepackage[makeroom]{cancel}
\usepackage{tikz}
\theoremstyle{definition}

\begin{document}

\begin{titlepage}

\begin{flushright}\vspace{-3cm}
{\small
IPM/P-2017/085 \\
\today }\end{flushright}
\vspace{0.5cm}

\begin{center}

{{\Large{\bf{Near-Horizon Extremal Geometries:}}}}\\
\vspace{2mm}
{{\Large{\bf{Coadjoint Orbits and Quantization}}}}\\
\vspace{12mm}
\centerline{\large{\bf{R. Javadinezhad\footnote{e-mail:rj1154@nyu.edu}$^{\dag}$, B. Oblak\footnote{e-mail:boblak@phys.ethz.ch}$^\flat$, M.M. Sheikh-Jabbari
\footnote{e-mail:jabbari@theory.ipm.ac.ir}$^\ddag$}}}
\vspace{8mm}
\normalsize
$^\dag$\textit{Physics Department, New York University, New York, USA}\\
\smallskip
$^\flat$\textit{Institut f\"ur Theoretische Physik, ETH Z\"urich, CH-8093 Z\"urich, Switzerland}\\
\smallskip
$^\ddag$\textit{School of Physics, Institute for Research in Fundamental Sciences (IPM), \\ P.O.Box 19395-5531, Tehran, Iran}\\
\smallskip


\vspace{5mm}

\begin{abstract}
\noindent The NHEG algebra is an extension of Virasoro introduced in [arXiv:1503.07861]; it describes the symplectic symmetries of $(n+4)$-dimensional Near Horizon Extremal Geometries with SL(2,$\mathbb{R})\times$U(1)$^{n+1}$ isometry. In this work we construct the NHEG group and classify the (coadjoint) orbits of its action on phase space. As we show, the group consists of maps from an $n$-torus to the Virasoro group, so its orbits are bundles of standard Virasoro coadjoint orbits over $T^{n}$. We also describe the unitary representations that are expected to follow from the quantization of these orbits, and display their characters. Along the way we show that the NHEG algebra can be built from u(1) currents using a twisted Sugawara construction.
\end{abstract}

\end{center}


\end{titlepage}
\setcounter{footnote}{0}
\renewcommand{\baselinestretch}{1.05}  

\setcounter{tocdepth}{1}
\tableofcontents
~
\hrule

\section{Introduction}

Since the seminal work of Bondi-van der Burg-Metzner, Sachs \cite{BMS}, and Brown-Henneaux \cite{Brown-Henneaux}, it is now an established fact that in generally covariant field theories there is a certain class of space-time diffeomorphisms (``residual'' or ``asymptotic'' symmetries) to which one can associate conserved surface charges. Therefore, certain sets of geometries are physically distinct despite being related by coordinate transformations. This statement can actually be adapted to any gauge theory \cite{Barnich-Brandt} and entails a refinement of the na\"ive notion of general covariance \cite{Residual-symmetry}. It is also consistent with holography \cite{AdS/CFT} since the conserved charges associated with asymptotic symmetries are fluxes of various field combinations at the space-time boundary \cite{Barnich-Brandt,CPSM,Lee-Wald,BMS-CFT,Thesis-Compere-Seraj-Hajian}.

One elegant application of asymptotic symmetries is the use of universal Cardy-like formulas \cite{Cardy-1986} to reproduce black hole entropy. This was first done in \cite{Strominger-1997} for BTZ black holes in AdS$_3$ \cite{BTZ} and was subsequently generalized to extremal Kerr black holes \cite{Kerr/CFT}, leading to the Kerr/CFT correspondence. In \cite{Kerr/CFT} the authors zoomed in on the near-horizon region of a four-dimensional extremal black hole with angular momentum $J$ and looked for asymptotic symmetries of the near-horizon region itself, finding a Virasoro algebra with central charge $c=12J$. Assuming that this reflects the presence of a (chiral) two-dimensional conformal field theory (CFT) and that the Cardy formula is applicable, they reproduced the entropy of the underlying extremal black hole. In this approach, both extremality and the near-horizon approximation are crucial. A similar analysis was subsequently applied to extremal black holes in various theories and dimensions, see \eg \cite{Compere-Kerr-CFT-review} and references therein. Later developments extended these results in two ways:
\begin{enumerate}
\item It was shown that the Virasoro algebra of Near Horizon Extreme Kerr consists of {\it symplectic symmetries}\footnote{The name ``symplectic symmetry'' was coined in \cite{Compere-symplectic} for cases where the pre-symplectic density $\omega$ vanishes. Symplectic (as opposed to asymptotic) symmetries are such that surface charges can be defined on any codimension two compact spacelike surface --- not necessarily at infinity \cite{CHSS-1, CHSS-2,CMSS, HS}. Apart from NHEGs, examples of symplectic symmetries include Brown-Henneaux transformations of Ba\~nados geometries \cite{Banados-geometries,CMSS,Afshar-et-al} and ADM charges \cite{HS}.}\cite{CHSS-1, CHSS-2} that do not actually require fall-off conditions for the metric; instead one can construct a family of mutually diffeomorphic but physically inequivalent solutions that span a phase space of ``boundary gravitons''.
\item The action of symplectic symmetry transformations on space-time can be extended beyond the near-horizon region, everywhere in the bulk of the extremal black hole \cite{Extreme-Kerr-Fluff}.
\end{enumerate}

In this work we study the symplectic symmetry group of the near-horizon region of a large class of extremal black holes in diverse dimensions, specifically the {\it Near-Horizon Extremal Geometries} (NHEGs) of \cite{CHSS-1,CHSS-2,Ishibashi-Hollands}. These are solutions of $(n+4)$-dimensional vacuum Einstein equations with SL$(2,\mathbb{R})\times$U$(1)^{n+1}$ isometry. The near-horizon geometries of extremal Kerr ($n=0$) and five-dimensional Myers-Perry black holes ($n=1$) fall in this class,\footnote{For $n=0$, near-horizon extreme Kerr is the unique metric in this class while for $n=1$ there are other solutions, \eg those obtained in the near horizon limit of extremal black rings or boosted Kerr strings \cite{5d-rings}.} while higher-dimensional Myers-Perry solutions ($n>1$) do not since they lack U$(1)^{n+1}$ isometries. In that setting, black hole entropy arises as a Noether charge that also happens to be the central charge of the symplectic symmetry algebra \cite{HSS,CHSS-2}.  It is subject to laws of NHEG mechanics that can be obtained from the zero temperature limit of standard black hole thermodynamics \cite{First-law}.

Accordingly, our motivation is to make progress towards the identification of the microstates responsible for NHEG entropy by classifying the possible homogeneous phase spaces with NHEG symmetry.  Since the NHEG group extends the Virasoro group familiar from two dimensional CFTs, our first goal will be to define it abstractly, including the central extension that plays a key role for entropy-matching. Similarly to the Virasoro group that extends the group Diff$(S^1)$ of diffeomorphisms of the circle, the NHEG group will be based on Diff$(T^{n+1})$, the group of diffeomorphisms of the torus, albeit with an anisotropy vector $\vec k$ related to the $n+1$ angular momenta of the background. (For $n=0$ the NHEG group reduces to the Virasoro group of Kerr/CFT.) It turns out that provided $\vk$ satisfies a natural quantization condition, the NHEG group is a bundle of Virasoro groups over an $n$-dimensional torus $T^{n}$. Equipped with these prerequisites we shall classify the orbits of NHEG backgrounds  under the NHEG group, \ie its coadjoint orbits; these come equipped with a natural symplectic form left invariant by the action of NHEG transformations. Owing to the similarity between NHEG and Virasoro algebras, this classification will be closely related to that of standard Virasoro orbits \cite{Lazutkin,KirillovVirasoro,Witten,Balog}. Each NHEG orbit can be seen as a set of physically inequivalent field configurations dressing a given background, and our goal is effectively to classify all possible such dressings. Note that, within a given orbit, all points correspond to space-time metrics with SL$(2,\mathbb{R})\times$U$(1)^{n+1}$ isometry, having identical $n+1$ angular momentum charges and identical entropy. We will also quantize the orbits\footnote{This procedure will rely on the unproven assumption that coadjoint orbits of the Virasoro group can be quantized, which is a thorny mathematical issue \cite{Salmasian:2014wwa}. We make no claims of rigour and confine the discussion to the formal level.} (promoting Poisson brackets to commutators) and build irreducible unitary representations of the NHEG algebra. These representations are continuous tensor products of Verma modules over $T^n$, which will allow us to evaluate their characters.

The plan is as follows. We start in section \ref{secNHEG} by reviewing the construction of NHEG symmetries in the gravitational context. This is then used in section \ref{secDef} to motivate an abstract definition for the NHEG group, its algebra and their central extensions. In section \ref{secOrb} we apply this definition to classify coadjoint orbits of the NHEG group. Section \ref{secSug} contains a free-field, twisted Sugawara construction of the NHEG algebra. In section \ref{Sec6} we describe unitary representations of this algebra and compute their characters. We end in section \ref{secCon} with some physical implications and applications.

\paragraph{Notation.} The algebra of vector fields on a circle (the Witt algebra) will be denoted as Vect$(S^1)$, and its central extension (Virasoro) as $\widehat{\text{Vect}}(S^1)$. The corresponding groups are Diff$(S^1)$ and $\widehat{\text{Diff}}(S^1)$, respectively. The NHEG algebra, the NHEG group and their central extensions will be respectively denoted by Vect$_{\vec k}(T^{n+1})$, $\NHEG$, $\widehat{\text{Vect}}{}_{\vec k}(T^{n+1})$ and $\widehat{\text{Diff}}{}_{\vec k}(T^{n+1})$.

\section{Near-horizon extremal geometries}
\label{secNHEG}

In this section we review the Near-Horizon Extremal Geometries studied in \cite{CHSS-1,CHSS-2,Ishibashi-Hollands}; they solve the $(n+4)$-dimensional vacuum Einstein equations and have an SL$(2,\mathbb{R})\times$U$(1)^{n+1}$ isometry group.

\paragraph{Metrics.} Following \cite{CHSS-2}, we consider an $(n+4)$-dimensional space-time endowed with a time coordinate $t$, a radial coordinate $r$, an azimuthal coordinate $\theta\in[0,\pi]$  and angular coordinates $\phi^i\in\mathbb{R}$ where $i=1,...,n+1$. These angles are identified as $\phi^i\sim\phi^i+2\pi$, so they label the points of an $(n+1)$-torus $T^{n+1}$. We shall think of the coordinates $(t,r,\theta,\phi^1,...,\phi^{n+1})$ as being defined in the near-horizon region of an extremal black hole; the metric of that region is fixed by SL(2,$\mathbb{R})\times$U(1)$^{n+1}$ isometry as
\be\label{NHEG-metric-generic}
ds^2= \Gamma(\theta)\left[-r^2dt^2+\frac{dr^2}{r^2}+d\theta^2+\gamma_{ij}(\theta)\big(d\phi^i-k^i rdt\big)\big(d\phi^j-k^j rdt\big)\right],
\ee
where the functions $\Gamma, \gamma_{ij}$ are constrained by Einstein's equations. They generally  depend on $(n+2)(n+1)/2$ parameters (integrals of motion), one of which is an overall normalization for the function $\Gamma$. Of the remaining parameters, $n$ are contained in the components ${k}^i$, which are arbitrary up to one relation, and the other $n(n+1)/2$ parameters fix the torus metric $\gamma_{ij}$ \cite{Ishibashi-Hollands}. The $n=0$  case corresponds to Near-Horizon Extreme Kerr \cite{BH,NHEK-uniqueness,Kerr/CFT} with $k^1=1$, while $n=1$ gives the near-horizon geometry of extremal Myers-Perry black holes or rings with $k^2=1/k^1$ \cite{BH,5d-rings}. Note that under SL$(2, \mathbb{R})$ isometries which keep the $t, r$ parts of the metric intact, $rdt$ transforms by a closed form, $rdt\to rdt + d\xi$, where $\xi$ depends on the details of the transformation. Hence SL$(2, \mathbb{R})$ isometries also involve translations of $\phi^i$ along $k^i$, $\phi^i\to \phi^i+k^i\xi$, so the vector $\vk$ also affects the generators (Killing vectors) of the SL(2,$\mathbb{R}$) isometry \cite{HSS}.

NHEGs are not black holes (they have no event horizon), but they do have infinitely many bifurcate Killing horizons at constant $t,r$, all at the same Frolov-Thorne \cite{Frolov-Thorne} temperature $1/2\pi$ \cite{HSS}; see \cite[sec.\ 2.1]{CHSS-2} for details. The metric on each bifurcation surface is
\be
\label{NHEG-horizon-metric}
ds^2_{\mathcal{H}}
=
\Gamma(\theta)\left(d\theta^2+\gamma_{ij}(\theta)d\phi^i d\phi^j\right).
\ee
It is smooth for all values of $\theta$; although $\vk$ does not appear here, it is a measurable physical parameter as its components are related to the angular momentum of the black hole --- see \cite{HSS, CHSS-2}. Thus, on the horizon, and of course on the whole NHEG \eqref{NHEG-metric-generic}, one is dealing with an \emph{anisotropic torus} --- a torus with a preferred direction specified by $\vk$. In principle, the components of $\vk$ may take any value; in practice however, the ratios of those components are directly related to ratios of components of angular momentum. Assuming that the latter is quantized, this implies that $\vk$ is proportional to a vector with integer entries, \ie an element of the dual lattice of $T^{n+1}$. Throughout this work we will always assume that this condition is satisfied, as it will be necessary to ensure smoothness of the NHEG group.

Let us then consider an anisotropic torus whose $\vk$ belongs to the dual lattice. In that case one can use the SL$(n+1,\mathbb{Z})$ volume-preserving symmetry of the torus to bring $\vk$, possibly up to normalization, to the convenient form
\be\label{kSim}
\vk=(0,...,0,1).
\ee
However, note that smoothness of the metric \eqref{NHEG-horizon-metric} generally prevents one from finding a global frame where $\vk$ has this form for all values of $\theta$. This fact will make the connection between NHEG orbits and the corresponding space-time metrics somewhat subtle; see section \ref{secCon}.

\paragraph{Symplectic symmetries.}  Consider a metric $g_{\mu\nu}$ solving Einstein's equations and let $\chi_1$, $\chi_2$ be vector fields; the corresponding symplectic density is a two-form in field space,  $\omega(\delta_{\chi_1}g_{\mu\nu}, \delta_{\chi_2}g_{\mu\nu}; g_{\mu\nu})$, where $\delta_\chi g_{\mu\nu}={\cal L}_\chi g_{\mu\nu}$. This density can be of the Lee-Wald \cite{Lee-Wald} or Barnich-Brandt \cite{Barnich-Brandt} type, or either of them up to a boundary term (see \cite{CHSS-2} for details). When $\omega$ vanishes on-shell for suitable vector fields, the latter generate \emph{symplectic symmetries} \cite{CHSS-2, CMSS, HS}. In such cases,  the integrability condition on charge variations is usually satisfied automatically and surface charges can be defined by integration over generic compact, space-like, codimension-two surfaces. In contrast to the perhaps more familiar asymptotic symmetries, these surfaces need not be at infinity; for the NHEG \eqref{NHEG-metric-generic} they can be located at arbitrary $(t,r)$. Furthermore, one can view the charges as generators of symplectomorphisms on a phase space built by acting on a background metric with finite diffeomorphisms generated by $\chi$'s. Each such phase space is an orbit of the symplectic symmetry group.

For the NHEG \eqref{NHEG-metric-generic}, an interesting family of vector fields generating symplectic symmetries is given by \cite{CHSS-1}
\be
\label{ASK}
\chi[{\epsilon}(\vec{\phi})]={\epsilon}\,\vec{k}\cdot\vec{\p}-\vec{k}\cdot\vec{\p}{\epsilon}\;\Big(\dfrac{1}{r}\p_{t}+r\p_{r}\Big),
\ee
where we write $\vec\phi=(\phi^1,...,\phi^{n+1})$ and $\eps=\eps(\vec\phi)$ is an arbitrary function on $T^{n+1}$ (it is $2\pi$-periodic in all $\phi^i$'s); we also let $\vec{\p}$ be the gradient operator $(\p_{\phi^1},...,\p_{\phi^{n+1}})$ on $T^{n+1}$ and write $\vk\cdot\vec\partial=k^i\partial_{\phi^i}$. These vector fields, unlike those of generic asymptotic symmetries, are exact in $r$.  By definition, their Lie brackets span the {\it NHEG algebra}; its structure is most easily described by defining generators $\chi_{\vec n}=\chi[e^{i\vec n\cdot\vec\phi}]$, $\vec n\in\mathbb{Z}^{n+1}$, whose brackets read
\be\label{bacha}
i[\chi_{\vec{m}},\chi_{\vec{n}}]
=
\vec{k}\cdot(\vec{m}-\vec{n})\,\chi_{\vec{m}+\vec{n}}.
\ee
It was shown in \cite{CHSS-1,CHSS-2} that conserved charges associated with the vector fields (\ref{ASK}) are well-defined and that they close according to the NHEG algebra up to a classical central extension. Indeed, if we denote by $L_{\vec n}$ the surface charge corresponding to $\chi_{\vec n}$, one has the Poisson brackets
\be
\label{sufacha}
i\{L_{\vec{m}}, L_{\vec{n}}\}
=
\vec{k} \cdot (\vec{m}- \vec{n}) L_{\vec{m}+\vec{n}}
+
\frac{c}{12} (\vec{k}\cdot \vec{m})^3\delta_{\vec{m}+\vec{n},0},
\ee
with
\be
\frac{c}{12}=\frac{S}{2\pi}=\vec{k}\cdot{\vec{J}},
\ee
where $S$ and $\vec{J}$  are respectively the entropy and angular momenta of the underlying extremal black hole, or equivalently of the corresponding near-horizon geometry.

Starting from a vector field (\ref{ASK}), one can exponentiate it to get a first glimpse of the (centrally extended) NHEG group. This exponential is a finite diffeomorphism $x\mapsto\bar x$ determined by the flow of $\chi[\epsilon]$ and it takes the form \cite{CHSS-2}
\be\label{finiteDiff}
\bar{\phi}^i=\phi^i + k^i F(\vec{\phi}),
\qquad
\bar{r} =re^{-{\Psi(\vec{\phi})}},
\qquad
\bar{t} =t-\frac{1}{r}(e^{\Psi(\vec{\phi})}-1),
\qquad
e^\Psi\equiv 1+\vk\cdot\vec{\p} F,
\ee
which indeed reduces to \eqref{ASK} for $F=\eps\ll 1$. (Note that we are not assuming anything about the norm of $\vk$.) The NHEG phase space is obtained by applying all diffeomorphisms of the form (\ref{finiteDiff}) to the background metric (\ref{NHEG-metric-generic}); it results in a family of metrics of the form
\be\label{g-F}
ds^2=\Gamma(\theta)\Big[-\left( \boldsymbol\sigma -  d \Psi \right)^2+\Big(\dfrac{dr}{r}-d{\Psi}\Big)^2
+d\theta^2+\gamma_{ij}(d\tilde{\phi}^i+{k^i}\boldsymbol{\sigma})(d\tilde{\phi}^j+{k^j}\boldsymbol{\sigma})\Big],
\ee
with
\be
\boldsymbol{\sigma}=e^{-{\Psi}}rdt+(1-e^{-\Psi})\dfrac{dr}{r},
\qquad
\tilde{\phi}^i=\phi^i+k^i (F-{\Psi}).
\ee
Even though each such metric is related to the background (\ref{NHEG-metric-generic}) by a diffeomorphism (\ref{finiteDiff}), the resulting space-time manifolds should be seen as genuinely distinct configurations of the gravitational field because their surface charges differ. Thus, starting from a given NHEG background, one obtains an entire family of physically distinct metrics labelled by different functions $F(\vec\phi)$; this family spans an orbit of the (centrally extended) NHEG group. One of the purposes of this paper is precisely to classify all such orbits and see how the metrics \eqref{g-F} fit in that classification.

\paragraph{NHEG charges.} On the phase space of metrics \eqref{g-F}, the surface charges $L_{\vec n}$ generating NHEG transformations as in (\ref{sufacha}) read \cite{CHSS-2}
\be
\label{charge}
L_{\vec{n}}= \int_\mathcal{H} \!\boldsymbol{\Omega}\ T[\Psi] e^{-i \vec{n}\cdot \vec{\phi} },
\ee
where $\mathcal H$ is the horizon with metric \eqref{NHEG-horizon-metric}, $\boldsymbol{\Omega}$ is its volume form  and $T$ is the ``stress tensor''
\be
\label{stress tensor}
T[\Psi]= \frac{1}{16\pi G}\Big((\Psi' )^2-2 \Psi''+2e^{2\Psi}\Big)
\ee
where primes denote directional derivatives along $\vec{k}$, \ie $\Psi' = \vec{k}\cdot \vec{\partial}\Psi$. Now, the transformation law of $\Psi$ (as defined in (\ref{finiteDiff})) under the NHEG algebra is \cite{CHSS-2}
\be
\delta_\eps\Psi= \eps\Psi'+\eps',\qquad \delta_\eps e^{\Psi}=(\eps e^\Psi)',
\ee
which is to say that $e^\Psi$ is a primary field with unit weight under the Virasoro transformations generated by $L_{\vn}$'s whose $\vn$ is proportional to $\vec{k}$. Consistently with this observation, the stress tensor \eqref{stress tensor} can be written in a more inspiring form. Indeed, defining a Schwarzian derivative
\be
\{{\cal F}(\vec\phi);\vec\phi\,\}\equiv \frac{{\cal F}'''}{{\cal F}'}-\frac{3}{2}\frac{{\cal F}''^2}{{\cal F}'^2},
\ee
one finds that for ${\cal F}'\equiv e^\Psi$ the expression (\ref{stress tensor}) can be recast as
\be
\label{Scharz transform}
T[\Psi]=\frac{1}{8\pi G}\big({\cal F}'^2-\{{\cal F};\vec{\phi}\}\big).
\ee
Here ${\cal F}$ is related to the $F$ of \eqref{finiteDiff} by ${\cal F}(\vec\phi)=\vk\cdot\vec{\phi}/|k|^2+F(\vec\phi)$. We will use this suggestive rewriting below to relate the family of metrics \eqref{g-F} to a NHEG coadjoint orbit.

\section{NHEG group and algebra}
\label{secDef}

In this section we provide an abstract definition of the NHEG group and its algebra, including central extensions. The main goal is to recover and extend the structures that emerge from the symmetry analysis summarized in the previous section. In particular we shall assume throughout that the anisotropy vector $\vec k$ takes the simple form (\ref{kSim}), which entails no loss of generality within the class of vectors $\vk$ whose components have rational ratios. The classification of the corresponding coadjoint orbits is postponed to section \ref{secOrb}.

\subsection{The NHEG group}

To define the NHEG group we proceed in two steps: first dealing with its centerless form, then adding central extensions.

\paragraph{Centerless version.} Consider an $(n+1)$-torus $T^{n+1}=\mathbb{R}^{n+1}/\mathbb{Z}^{n+1}$ with coordinates $\phi^1,...,\phi^{n+1}\in\mathbb{R}$, each identified as $\phi^i\sim\phi^i+2\pi$. The NHEG transformations generated by vector fields of the form (\ref{ASK}) act on that torus according to
\be
\vec\phi\mapsto\vec\phi+\epsilon(\vec\phi)\vec k.
\label{s1}
\ee
The set of such infinitesimal transformations is a subalgebra of Vect$(T^{n+1})$ that we denote by Vect$_{\vk}(T^{n+1})$.  For $\vk=(0,...,0,1)$, eq.\ (\ref{s1}) reduces to $\phi^{n+1}\mapsto\phi^{n+1}+\epsilon(\vec\phi)$, with $\phi^1,...,\phi^n$ left unchanged; the exponential of this transformation is a diffeomorphism
\be
(\phi^1,...,\phi^n,\phi^{n+1})\mapsto\big(\phi^1,...,\phi^n,\cF(\phi^1,...,\phi^{n+1})\big)
\label{helsinki}
\ee
where $\cF$ would have been written as $\cF(\vec\phi)=\phi^{n+1}+F(\vec\phi)$ with the notation of  (\ref{finiteDiff}). It is such that
\be
\begin{split}
\cF(\phi^1,...,\phi^n,\phi^{n+1}+2\pi) &= \cF(\phi^1,...,\phi^n,\phi^{n+1})\pm2\pi\quad\text{and}\quad\partial\cF/\partial\phi^{n+1}\neq0,\\
\cF(\phi^1,...,\phi^i+2\pi,...,\phi^{n+1}) &= \cF(\phi^1,...,\phi^n,\phi^{n+1})+2\pi N_i\;\;\forall\,i=1,...,n
\end{split}
\label{t3}
\ee
where $N_1,...,N_n$ are integers that may take different values for different $\cF$'s (so the last line is just the requirement that $\cF$ be $2\pi$-periodic in $\phi^1,...,\phi^n$ modulo $2\pi$). Thus the NHEG group is
\be
\NHEG
=
C^{\infty}\big(T^{n},\text{Diff}(S^1)\big).
\label{ss3}
\ee
It is the set of smooth maps that send a point $(\phi^1,...,\phi^n)$ on a circle diffeomorphism $\cF(\phi^1,...,\phi^n, \cdot)$. In other words, it is a bundle of Diff$(S^1)$'s over $T^n$, which already suggests that its central extension will be a bundle of Virasoro groups over $T^n$. All our later observations follow from this basic fact.

To lighten the notation, from now on we write $\phi^{n+1}\equiv\phii$ and $(\phi^1,...,\phi^n)\equiv\Phi$, as well as
\be
\cF(\phi^1,...,\phi^n, \phi^{n+1})
=
\cF(\phii,\Phi)
\equiv
\cF_{\Phi}(\phii)
\ee
so that $\cF(\phi^1,...,\phi^n,\cdot)=\cF_{\Phi}$. We also denote partial derivatives with respect to $\phi^{n+1}=\phii$ by a prime: $\partial\cF/\partial\phi^{n+1}=\cF'$. Derivatives with respect to the remaining coordinates $\phi^1,...,\phi^n$ will never appear, as these angles are mere spectators or ``parameters'' from the point of view of the NHEG group. In particular, the group operation is $(\cF,\cG)\mapsto\cF\cdot\cG$ with
\be
(\cF\cdot\cG)(\phii,\Phi)
=
\cF\big(\cG(\phii,\Phi),\Phi\big),
\quad
\text{\ie}
\quad
(\cF\cdot\cG)_{\Phi}
=
\cF_{\Phi}\circ\cG_{\Phi}
\;\;\quad \forall\,\Phi\in T^{n}.
\label{grop}
\ee
Note that for $n=0$ the ``torus'' $T^{n}=T^0$ contains only one point and the NHEG group (\ref{ss3}) reduces to $\text{Diff}(S^1)$.

A remark: since the fundamental group of the torus $T^n$ is $\mathbb{Z}^n$, the NHEG group (\ref{ss3}) has infinitely many connected components (for $n>0$). Indeed, the second condition in (\ref{t3}) leaves room for winding numbers $N_i$; any two elements of the NHEG group whose winding numbers differ belong to different connected components. Furthermore, the group $\text{Diff}(S^1)$ has two connected components, corresponding to diffeomorphisms that preserve or break the orientation of the circle. In other words, each connected component of the NHEG group can be labelled by (i) the winding number of its elements, and (ii) the sign plus or minus in the first line of (\ref{t3}). In practice however, we will only deal with the maximal connected subgroup of the NHEG group, \ie the component of the identity. It consists of diffeomorphisms $\cF$ with zero winding number and which preserve the orientation of the circle so that $\cF'>0$. From now on we simply refer to this connected group as ``the NHEG group''.

\paragraph{Centrally extended version.} In order to reproduce the centrally extended surface charge algebra (\ref{sufacha}), we need to define a centrally extended version of the NHEG group. Since the coordinates $\phi^1,...,\phi^n$ behave as parameters, we can define a ``local'' central extension such that the corresponding central charge is a function of $\phi^1,...,\phi^n$. To see this, consider the set of all pairs $(\cF,\alpha)$ whose entries are (i) a diffeomorphism $\cF(\phi^1,...,\phi^n, \phi^{n+1})=\cF(\phii,\Phi)$ belonging to the NHEG group (\ref{ss3}), and (ii) a function $\alpha(\phi^1,...,\phi^n)=\alpha(\Phi)$ on the torus $T^{n}$. Consider then the group operation
\be
(\cF,\alpha)\cdot(\cG,\beta)
=
\big(
\cF\cdot\cG,
\alpha+\beta+C[\cF,\cG]
\big)
\label{s9}
\ee
where the function $C[\cF,\cG]$ on $T^{n}$ is a simple generalization of the Bott(-Thurston) cocycle \cite{bott1977characteristic}:
\be
C[\cF,\cG](\Phi)
=
-\frac{1}{48\pi}
\int_0^{2\pi}\!\!d\phii\,\log(\cF_{\Phi}'\circ\cG_{\Phi})\,\frac{\cG_{\Phi}''}{\cG_{\Phi}'}.
\ee
The set of such pairs $(\cF,\alpha)$ spans the centrally extended NHEG group,
\be
\label{ss3bis}
\widehat{\text{Diff}}{}_{\vk}(T^{n+1})
=
C^{\infty}\big(T^{n},\widehat{\text{Diff}}(S^1)\big);
\ee
it is an extension of (\ref{ss3}) by the space $C^{\infty}(T^{n})$ of smooth, real functions $\alpha$. This extension is central since it commutes with everything, and it implies that the NHEG algebra can have infinitely many central charges. The constant, $\Phi$-independent central charge of the surface charge algebra (\ref{sufacha}) is recovered upon replacing the space $C^{\infty}(T^{n})$ by $\mathbb{R}$ and replacing the group operation (\ref{s9}) by
\be
(\cF,\alpha)\cdot(\cG,\beta)
\equiv
\Big(
\cF\cdot\cG,
\alpha+\beta+
\int_{T^{n}}\!\!d\Phi\,C[\cF,\cG]
\Big),
\label{s9b}
\ee
where now $\alpha,\beta\in\mathbb{R}$ and  $d\Phi\equiv d\phi^1...d\phi^n/(2\pi)^n$ so that $\int_{T^n}\!\!d\Phi=1$. In other words, the central extension here is the zero-mode projection of the $\Phi$-dependent one in (\ref{s9}). As we now verify, the Lie algebra that follows from these constructions contains and extends (\ref{sufacha}).

\subsection{The NHEG algebra}

Eq.\ (\ref{s1}) says that the Lie algebra of the NHEG group consists of vector fields $\bfeps=\eps(\vec\phi)\vec k\cdot\vec\p$ where $\vec\p$ is the gradient on $T^{n+1}$. With the convention $\vk=(0,...,0,1)$, any such vector field takes the form $\bfeps=\eps(\phii,\Phi)\partial_{\phii}\equiv \eps_{\Phi}(\phii)\partial_{\phii}$, so from now on we write any element of the NHEG algebra as $\bfeps$ or $\eps_{\Phi}$. To obtain the adjoint representation of the NHEG group, we use the general definition
\be
\big(\text{Ad}_{\cF}\bfeps\big)(\phii,\Phi)
\equiv
\left.\frac{d}{dt}\right|_{t=0}\Big(\cF\cdot e^{t\bfeps}\cdot\cF^{-1}\Big)(\phii,\Phi)
\ee
where the product of group elements is (\ref{grop}), so it is a $\Phi$-pointwise multiplication on $T^{n}$ given by composition along the $S^1$ spanned by $\phii$. Thus the computation is the same as for the Virasoro group (see \eg \cite[sec.\ 4.4.2]{Guieu} or \cite[sec.\ 6.1.4]{Blagoje-thesis}), up to an extra parametric dependence on $\Phi\in T^{n}$. The result is
\be
\big(\text{Ad}_{\cF}\bfeps\big)_{\Phi}(\phii)
=
\frac{\eps_{\Phi}(\cF_{\Phi}^{-1}(\phii))}{(\cF_{\Phi}^{-1})'(\phii)},
\quad
\text{\ie}
\quad
\big(\text{Ad}_{\cF}\bfeps\big)_{\Phi}(\cF_{\Phi}(\phii))
=
\cF_{\Phi}'(\phii)\eps_{\Phi}(\phii).
\ee
We can repeat the same arguments with the centrally extended group (\ref{s9}); once more the result is the same as in the Virasoro case up to a parametric dependence on $\Phi$:
\be
\widehat{\text{Ad}}{}_{(\cF,\alpha)}(\bfeps,\beta)
=
\Big(
\text{Ad}_{\cF}\bfeps,
\beta-\frac{1}{24\pi}\int_0^{2\pi}\!\!d\phii\,\eps(\phii)\{\cF;\phii\}
\Big),
\label{ad}
\ee
where $\alpha$, $\beta$ and the second entry on the right-hand side are functions of $\Phi\in T^{n}$. For instance, what we denote by $\{\cF;\phii\}$ is the function of $\Phi$ and $\phii$ such that $\{\cF;\phii\}(\Phi)\equiv\{\cF_{\Phi};\phii\}$, with $\{\cF;\phii\}=\cF'''/\cF'-\tfrac{3}{2}(\cF''/\cF')^2$ the standard Schwarzian derivative.

The adjoint representation yields the Lie bracket of the algebra thanks to the definition
\be
\big[(\bfeps,\alpha),(\boldsymbol{\zeta},\beta)\big]
\equiv
-\left.\frac{d}{dt}\right|_{t=0}\widehat{\text{Ad}}{}_{(e^{t\bfeps},t\alpha)}(\boldsymbol{\zeta},\beta),
\label{libra}
\ee
where the minus sign is a matter of convention. (Note that we are working with the centrally extended group.) Using (\ref{ad}) and evaluating the $t$ derivative, one finds
\be
\big[(\bfeps,\alpha),(\boldsymbol{\zeta},\beta)\big]
=
\Big(
[\bfeps,\boldsymbol{\zeta}],
\frac{1}{24\pi}\int_0^{2\pi}\!\!d\phii\,\eps'''(\phii)\zeta(\phii)
\Big),
\label{libasi}
\ee
where, as before, $\alpha$, $\beta$ and the second entry on the right-hand side are functions of $\Phi$. The first entry of the right-hand side involves the standard Lie bracket of vector fields. To make the structure of the algebra more explicit, we can expand all functions  in Fourier modes on the torus. Thus, we let $\vec m,\vec n\in\mathbb{Z}^{n+1}$ and define the NHEG generators
\be
L_{\vec m}\equiv\big(e^{i\vec m\cdot\vec\phi}\vk\cdot\vec{\partial},0\big).
\ee
We also take $\vec M\in\mathbb{Z}^{n}$ and define the (infinitely many) central charges
\be\label{central-charge}
c_{(\vec M, 0)}\equiv(0, e^{i\vec M\cdot\Phi}).
\ee
Then the Lie bracket (\ref{libasi}) implies that the central charges commute with everything, while the brackets of NHEG generators lead to the expected centrally extended NHEG algebra:
\be
i\big[L_{\vec m},L_{\vec n}\big]
=
\vec k\cdot(\vec m-\vec n)L_{\vec m+\vec n}
+
\frac{c_{\vec m+\vec n}}{12}(\vec k\cdot\vec m)^3\delta_{\vec k\cdot(\vec m+\vec n),0},
\label{NHEG-algebra}
\ee
where $\vec k\equiv(0,...,0,1)$. The same formula would hold for arbitrary $\vk$ on the dual lattice of $T^{n+1}$. Note that this is the unique maximal central extension of the centerless NHEG algebra (up to redefinitions of generators, such as shifts in $L_{\vec 0}$); this can be verified directly by requiring the central terms to satisfy Jacobi identities. In particular, there is no Kac-Moody central extension of the type that might be expected for maps from $T^n$ to a generic Lie algebra; the reason is that the Virasoro algebra has no invariant bilinear form.

The centerless version of the bracket (\ref{NHEG-algebra}) clearly reproduces the symplectic symmetry algebra (\ref{bacha}), while the restriction of central charges to their zero-mode reproduces the surface charge algebra (\ref{sufacha}). Note that (\ref{NHEG-algebra}) has a Virasoro subalgebra generated by $L_{m\vk}$, and that the generators $L_{\vm}, \vm\cdot \vk=0$ span an Abelian subalgebra that may be viewed as $C^{\infty}(T^{n})$.  An equivalent way to phrase this is to introduce NHEG generating fields and their Fourier modes along $\vk$,
\be\label{L-varphi-Phi}
{\cal L}(\phii,\Phi)=\sum_{\vn}\ L_{\vn} e^{i\vn\cdot \vec{\phi}},
\qquad
L_{n}(\Phi)\equiv\frac{1}{2\pi}\int_0^{2\pi}\!\!d\phii\,{\cal L}(\phii,\Phi) e^{-in\phii}.
\ee
In these terms the bracket (\ref{NHEG-algebra}) reads
\be\label{Vir-bundle-algebra}
i\big[L_m(\Phi), L_n(\Phi')\big]
=
\left((m-n) L_{n+m}(\Phi) +\frac{c(\Phi)}{12}m^3\delta_{n+m,0}\right)\delta^{n}(\Phi-\Phi'),
\ee
where the central charges of \eqref{central-charge} are the Fourier modes of $c(\Phi)$. Thus the NHEG algebra is a bundle of Virasoro algebras over an $n$-dimensional torus $T^{n}$; this was of course expected from (\ref{ss3})-(\ref{ss3bis}). The remainder of this paper is devoted to some consequences of this observation.

\section{NHEG orbits}
\label{secOrb}

Consider a space-time metric that solves Einstein's vacuum equations and that takes the NHEG form (\ref{NHEG-metric-generic}). This metric is a point in an infinite-dimensional phase space that consists of gravitational fields representing the near-horizon geometry of an extremal black hole. The NHEG symmetry group acts on this phase space and relates various physically distinct metrics to one another; metrics that are related by NHEG transformations belong, by definition, to the same NHEG orbit. For instance, the metrics \eqref{g-F} span one such orbit.  As it turns out, the transformation law of NHEG metrics under the NHEG group is given by the coadjoint representation, so the purpose of this section is to study and classify NHEG coadjoint orbits. The result will be closely related to the classification of Virasoro orbits \cite{Lazutkin,KirillovVirasoro,Witten}; see \eg \cite{Balog} or \cite[chap.\ 7]{Blagoje-thesis} for a review.

By definition, any Lie group acts on its Lie algebra according to the adjoint representation, and the corresponding dual action is the coadjoint representation. In the present case, assuming $\vk=(0,...,0,1)$, the dual of the (centrally extended) NHEG algebra consists of pairs $(\bL,c)$ where $\bL=L_{\Phi}(\phii)d\phii^2$ is a $\Phi$-dependent Virasoro coadjoint vector while $c=c({\Phi})$ is a $\Phi$-dependent Virasoro central charge.\footnote{Strictly speaking, both $\bL$ and $c$ are the coefficients of volume forms $\bL\otimes d\Phi$ and $c\,d\Phi$ on $T^n$, but since NHEG transformations never affect the $\Phi$ coordinates this abuse of notation is harmless.} The pairing between $(\bL,c)$ and the NHEG algebra is
\be
\big<(\bL,c),(\bfeps,\alpha)\big>
=
\frac{1}{2\pi}\int_0^{2\pi}\!\!d\phii\int_{T^{n}} \!\!d\Phi\,L(\phii,\Phi)\epsilon(\phii,\Phi)
+
\int_{T^{n}} \!\!d\Phi\,c(\Phi)\alpha(\Phi).
\ee
For the constant central extension defined in (\ref{s9b}), the second term of the right-hand side reduces to a product $c\cdot\alpha$, where both $c$ and $\alpha$ are just real numbers.

Since the NHEG group is a bundle of Virasoro groups, its coadjoint representation coincides with the standard transformation law of (chiral) CFT stress tensors --- \ie the coadjoint representation of the Virasoro group ---, up to a parametric dependence on $\Phi$. Explicitly, this representation is defined by the general formula
\be
\big<\widehat{\text{Ad}}{}^*_{(\cF,\alpha)}(\bL,c),(\bfeps,\alpha)\big>
=
\big<(\bL,c),\widehat{\text{Ad}}{}_{(\cF,\alpha)^{-1}}(\bfeps,\alpha)\big>.
\ee
Using the fact that the NHEG adjoint action is (\ref{ad}), one finds
\be
\widehat{\text{Ad}}{}^*_{(\cF,\alpha)}(\bL,c)
=
\Big(
\text{Ad}{}^*_{\cF}\bL-\frac{c}{12}\{\cF^{-1};\cdot\},c
\Big),
\ee
where
\be
\Big(\text{Ad}{}^*_{\cF}\bL-\frac{c}{12}\{\cF^{-1};\cdot\}\Big)(\phii,\Phi)
=
\big[(\cF_{\Phi}^{-1})'(\phii)\big]^2L_{\Phi}(\cF_{\Phi}^{-1}(\phii))
-
\frac{c({\Phi})}{12}\{\cF_{\Phi}^{-1};\phii\}.
\label{cotra}
\ee
The same formula follows from the `rigid' central extension of (\ref{s9b}), except that the corresponding central charge is independent of $\Phi$; note the similarity with (\ref{Scharz transform}).

\begin{figure}[t]
\begin{minipage}{0.2\textwidth}
\centering
~
\end{minipage}\hfill
\centering
\begin{minipage}{0.3\textwidth}
\centering
\includegraphics[width=\linewidth]{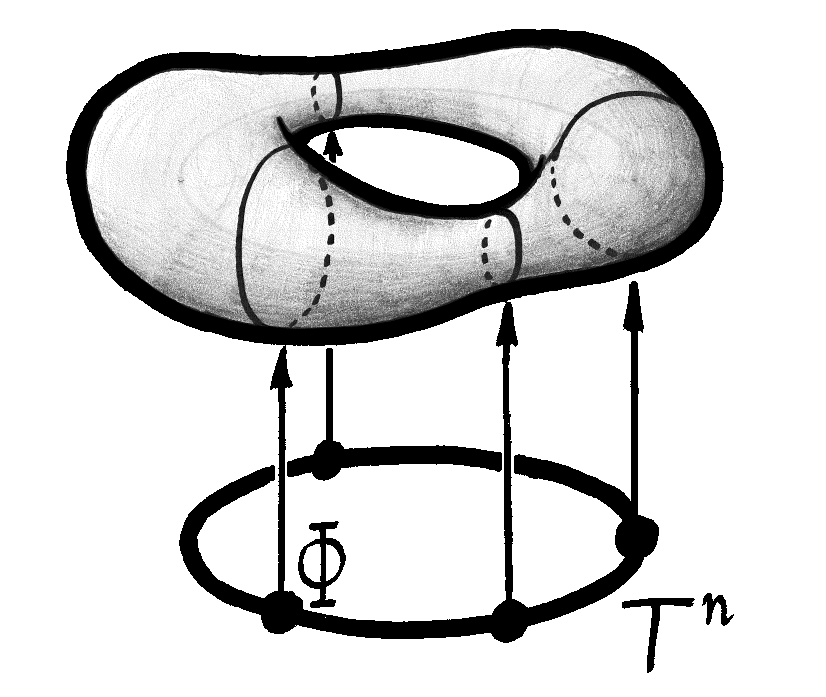}\par
(a)
\end{minipage}\hfill
\begin{minipage}{0.3\textwidth}
\centering
\includegraphics[width=\linewidth]{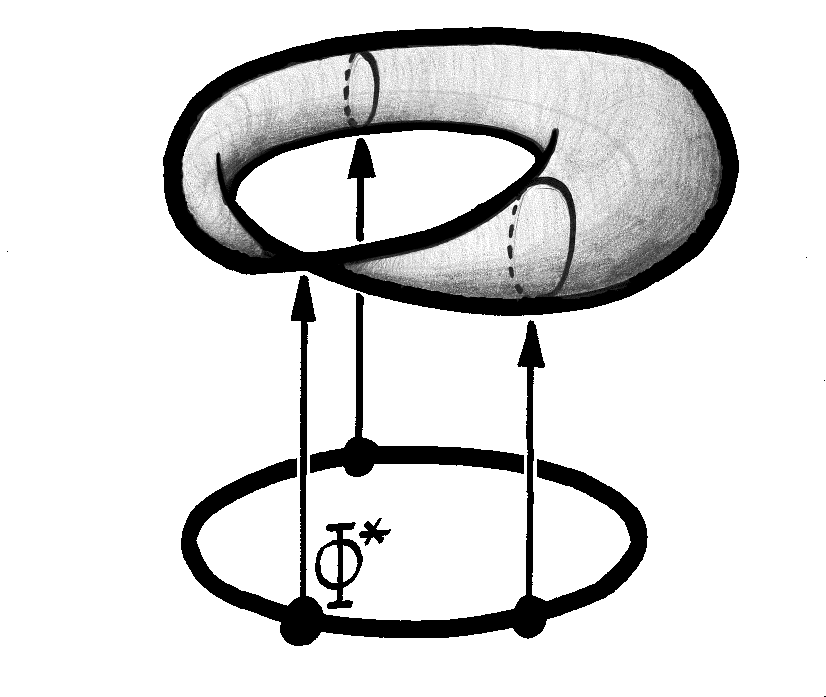}\par
(b)
\end{minipage}\hfill
\begin{minipage}{0.2\textwidth}
\centering
~
\end{minipage}
\caption{\label{FIG} Two coadjoint orbits of the NHEG group. Each is a fibre bundle of Virasoro orbits over $T^n$; for simplicity we take $n=1$ and represent Virasoro orbits by disks, so that NHEG orbits have the topology of solid tori. The radius of the disk varies with its position $\Phi$, reflecting the fact that different fibres generally correspond to different Virasoro orbits. In (a) the NHEG orbit is smooth, but in (b) the fibre at $\Phi^*$ is degenerate (it has zero radius) while its immediate neighbours are not, so the corresponding NHEG orbit is pathological. This is due to discontinuous jumps between Virasoro orbits of different types; it occurs for instance when a NHEG coadjoint vector $\bL$ crosses the critical value $L_{\Phi^*}=-c(\Phi^*)/24$. We discard such singular NHEG orbits from our classification and only include smooth orbits such as the one in (a).}
\end{figure}

The transformation law (\ref{cotra}) implies that each coadjoint orbit of the NHEG group is a bundle of Virasoro orbits over $T^{n}$; the fibre at $\Phi$ is the Virasoro orbit of the CFT stress tensor $L_{\Phi}$ with central charge $c({\Phi})$ --- see fig.\ \ref{FIG}. Schematically, any NHEG orbit $\cO$ takes the form of a disjoint union
\be
\label{Obobo}
\cO
=
\bigsqcup_{\Phi\in T^n}\cO_\Phi
\ee
where each $\cO_{\Phi}$ is a coadjoint orbit of the Virasoro group. With the $\Phi$-dependent central extension of (\ref{s9}), both the coadjoint vectors and the central charges of these Virasoro orbits generally vary with $\Phi$; by contrast, with the rigid central extension of (\ref{s9b}), the central charge takes the same value for all points $\Phi$ on $T^{n}$, but the stress tensors $L_{\Phi}$ vary. In principle, this achieves our goal: since the classification of Virasoro orbits is known, we can describe any NHEG orbit as a bundle of Virasoro orbits over $T^{n}$. For instance, the orbit (\ref{g-F}) is obtained when $c({\Phi})=6S/\pi$ and $L_{\Phi}=c/12$ are constant, with the identification between $\bL$ and the stress tensor of (\ref{Scharz transform}) given by $A\cdot T=\widehat{\text{Ad}}{}^*_{\cF^{-1}}L$ in terms of the horizon area $A=4GS$.

Note that, within a NHEG orbit, one can move between different Virasoro orbits by varying $\Phi$, but the orbits that are spanned in the process are not entirely arbitrary. Indeed, in order for the NHEG orbit to be smooth, there must be no singularity in the bundle (\ref{Obobo}); to ensure this, all $\cO_{\Phi}$'s should be specified by the same winding number and have the same type of  SL$(2,\mathbb{R})$ monodromies, meaning that the monodromy matrices of the associated Hill's equations should either be all elliptic, or all hyperbolic, or all mutually conjugate and parabolic;\footnote{Hill's equation and the associated monodromies and winding numbers are tools that are routinely used to classify Virasoro orbits \cite{Lazutkin,KirillovVirasoro}. See \eg \cite{Balog} or \cite[chap.\ 7]{Blagoje-thesis} for a review; see also eq.\ (\ref{Hill-equation}) below.} see again fig.\ \ref{FIG}. Thus, for example, if $L_{\Phi_0}=-c(\Phi_0)/24$ at some point $\Phi_0$, then the corresponding Virasoro orbit $\cO_{\Phi_0}$ has unit winding number and trivial monodromy up to a sign --- it is the orbit of a CFT vacuum stress tensor under conformal transformations ---; but smoothness then requires that any other $\cO_{\Phi}$ also be a Virasoro vacuum orbit, which means that there exists a NHEG transformation which turns the NHEG coadjoint vector $\bL$ into $\big(\cF\cdot\bL\big)(\phii,\Phi)=-c(\Phi)/24$ for any $\Phi\in T^n$. In other words, once the central charges $c(\Phi)$ are fixed, there exists a unique smooth NHEG orbit that contains a fibre which is a vacuum Virasoro orbit. Amusingly, this means that local data (from the point of view of $T^n$) contains global information within the class of smooth NHEG orbits: knowing $\cO_{\Phi_0}$ gives information on the other Virasoro orbits $\cO_{\Phi}$. Below we provide examples of smooth orbits for which the fibres at different $\Phi$'s are genuinely different Virasoro orbits; but first let us revisit the classification of NHEG orbits from a slightly different point of view.

\paragraph{Lie-algebraic data.} Following \cite{Witten}, a quick way to guess the classification of coadjoint orbits is to find the Lie algebras of their stabilizers. The starting point is the Lie algebra representation that corresponds to the coadjoint transformation law (\ref{cotra}). Taking $\cF(\phii,\Phi)=\phii+\eps(\phii,\Phi)$ in the latter equation and working to first order in $\eps$, one finds
\be
\text{ad}^*_{\bfeps}\bL
=
\eps(\phii,\Phi) L_{\Phi}'+2\eps(\phii,\Phi)' L_{\Phi}-\frac{c(\Phi)}{12}L_{\Phi}'''.
\label{epsil}
\ee
Now, the stabilizer of $\bL$ consists of those NHEG group elements that leave it fixed; in Lie-algebraic terms this is to say that the algebra of the stabilizer is spanned by $\bfeps$'s for which (\ref{epsil}) vanishes:
\be
\bfeps\in\text{stabilizer algebra}
\qquad\Longleftrightarrow\qquad
\eps(\phii,\Phi) L_{\Phi}'+2\eps(\phii,\Phi)' L_{\Phi}-\frac{c(\Phi)}{12}L_{\Phi}'''=0.
\ee
Of course, this equation and (\ref{epsil}) coincide with standard Virasoro expressions up to a parametric dependence on the transverse coordinates $\Phi$.

Knowing the stabilizer (call it $G_{\bL}$) of a NHEG coadjoint vector $\bL$, one infers that the corresponding orbit is diffeomorphic to a quotient of the NHEG group by $G_{\bL}$. For example, suppose that $c(\Phi)=c>0$ and $L_{\Phi}=L>-c/24$ are positive constants; then the stabilizer of $(\bL,c)$ is generated by all $\bfeps$'s such that $\eps'=0$, so it consists of maps from $T^{n}$ to the group U$(1)\cong S^1$ of rigid rotations. As a result the corresponding NHEG orbit is diffeomorphic to the quotient space
\be
{\cal O}_{(L,c)}
\cong
C^{\infty}\big(T^{n},\text{Diff}(S^1)\big)\Big/C^{\infty}\big(T^{n},S^1\big)
\cong
C^{\infty}\big(T^{n},\text{Diff}(S^1)/S^1\big)
\qquad
\text{for constant $L\neq-c/24$.}
\label{orexam}
\ee
Similarly, the stabilizer of $L_{\Phi}(\phii)=-c(\Phi)/24$ is the group of smooth maps from $T^{n}$ to PSL$(2,\mathbb{R})$. Note once more that smoothness of the orbit restricts the allowed configurations $L_{\Phi}(\phii)$: if $L_{\Phi_0}=-c(\Phi_0)/24$ at some point $\Phi_0$, then $G_{\bL}$ contains maps that send $\Phi_0$ on PSL$(2,\mathbb{R})$; but in order for $G_{\bL}$ to be smooth, its elements must be able to map any other point $\Phi$ on PSL$(2,\mathbb{R})$, which implies as before that $L_{\Phi}=-c(\Phi)/24$ for any $\Phi$.

\paragraph{Parametric Hill's equations.} The standard classification of Virasoro coadjoint orbits relies on the monodromies of Hill's equation \cite{Lazutkin}, so let us comment on the use of this method in the present context. (For a pedagogical review, see \cite{Balog} or \cite[chap.\ 7]{Blagoje-thesis}.) The starting point is a function $\psi(\vec\phi)$ that we take to be single-valued on $T^{n}$ (so it is $2\pi$-periodic in the angular coordinates contained in $\Phi$), but {\it not} necessarily periodic in $\phii$. Given a NHEG coadjoint vector $(\bL,c)$, we require $\psi$ to solve the parametric Hill's equation
\be
\label{Hill-equation}
-\frac{c(\Phi)}{6}\psi''(\vec\phi)+L_{\Phi}(\phii)\psi(\vec\phi)
=
0.
\ee
The same equation would be relevant to the classification of Virasoro orbits, albeit without parametric dependence on $\Phi$. The fact that $L_{\Phi}(\phii)$ is $2\pi$-periodic in $\phii$ implies that, for any $\Phi\in T^{n}$, any two linearly independent solutions $\psi_1,\psi_2$ of (\ref{Hill-equation}) behave in a quasi-periodic way:
\be
\begin{pmatrix} \psi_1(\phii+2\pi,\Phi) \\ \psi_2(\phii+2\pi,\Phi) \end{pmatrix}
=
\mathbf{M}(\Phi)
\begin{pmatrix} \psi_1(\phii,\Phi) \\ \psi_2(\phii,\Phi) \end{pmatrix},
\ee
where the monodromy matrix $\mathbf{M}(\Phi)$ belongs to SL$(2,\mathbb{R})$. Again, the only difference between this setting and the standard Virasoro case is the dependence on $\Phi$. In particular, similarly to the  Virasoro case, the conjugacy class of the monodromy matrix is an invariant label for the orbit of $(\bL,c)$; in the case at hand, this label is a function on $T^{n}$ since the conjugacy class of $\mathbf{M}(\Phi)$ depends on $\Phi$. Thus, each NHEG orbit is labelled by $\Phi$-dependent conjugacy classes in SL$(2,\mathbb{R})$. In order for the orbit to be smooth, these conjugacy classes must all be of the same type at all $\Phi$'s --- elliptic, hyperbolic or parabolic. In addition, Virasoro orbits are labelled by a discrete winding number \cite{Balog}; smooth NHEG orbits are such that this number is the same for all $\Phi\in T^{n}$.

We conclude with an example. Consider a NHEG coadjoint vector $L_{\Phi}(\phii)=L_{\Phi}$ that does not depend on $\phii$ (at fixed $\Phi$, it is a constant from the point of view of the coordinate $\phii$). One readily verifies that a corresponding normalized pair of solutions of Hill's equation (\ref{Hill-equation}) is
\be
\psi_1(\phii,\Phi)
=
\frac{\rho(\Phi)}{\sqrt{2K(\Phi)} }e^{K(\Phi)\phii},
\qquad
\psi_2(\phii,\Phi)
=
\frac{1}{\rho(\Phi)\sqrt{2K(\Phi)}} e^{-K(\Phi)\phii},
\qquad
L_\Phi(\phii)=\frac{c(\Phi)}{6}K^2(\Phi),
\label{exeppe}
\ee
where $\rho(\Phi)$ is any non-zero function on $T^{n}$. When the function $K(\Phi)$ is real, the corresponding Virasoro orbits are hyperbolic; for $K(\Phi)=i\nu(\Phi)/2$, $\nu\in(0,1)$, they are elliptic; and for $K(\Phi)=i/2$ they are vacuum orbits, for which $L_{\Phi}=-c(\Phi)/24$ and the monodromy matrix is minus the identity. In the first two cases the stabilizer consists of maps from $T^{n}$ to U$(1)$, as in (\ref{orexam}); in the vacuum case it is spanned by maps from $T^{n}$ to PSL$(2,\mathbb{R})$.

\newpage
\section{NHEG-Kac-Moody algebra}
\label{secSug}

In  this section we describe a ``free-field'' construction of the NHEG algebra in terms of Abelian current algebras. This is partly motivated by representation theory, but it is also of interest in computations of black hole entropy from near-horizon symmetries \cite{Afshar-et-al,Extreme-Kerr-Fluff}; we shall briefly return to the latter issue in the last section of this work.

Let us consider a set of currents $J_i(\vec{\phi}), \; i=1,...,n+1$ and assume that their Fourier modes
\be
\label{J-i-n}
J_{i,\vn}
=
\frac{1}{2\pi}\int_0^{2\pi}\!\!d\phii\int_{T^n}\!\!d\Phi\,e^{-i\vn\cdot\vec{\phi}}J_i(\vec{\phi})
\ee
satisfy the centrally extended algebra
\be
\label{J-i-n-parallel-perp}
i[J_{i,\vm},J_{j,\vn}]
=
(\vk\cdot\vm)\, g_{ij}\,\de_{\vm+\vn,0}\,Z
\ee	
where $Z$ is a central charge and $g_{ij}$ is a {constant} metric on the torus $T^{n+1}$. This metric is arbitrary and none of our results will depend on it, so we may choose it to be the same as in (\ref{NHEG-metric-generic}) at some given $\theta$. We refer to (\ref{J-i-n-parallel-perp}) as the {\it NHEG-Kac-Moody} algebra.

When $\vk$ is proportional to a vector in the dual lattice of $T^{n+1}$ (which we assume to be the case),  we can take $\vk=(0,...,0,1)$ and use the notation of section \ref{secDef}. Then eq.\ (\ref{J-i-n-parallel-perp}) can be written as
\be
\label{J-Phi-algebra}
J_{i,m}(\Phi)
=
\frac{1}{2\pi} \int_0^{2\pi}\!\!d\phii\,J_i(\vec\phi) e^{im\phii},
\qquad
i[J_{i,m}(\Phi), J_{j,n}(\Phi')]= m\, g_{ij}\,\de_{m+n,0}\ \delta^{n}(\Phi-\Phi').
\ee
As we see, $J_{i,0}(\Phi)$ is central for all $\Phi\in T^{n}$. To recover the NHEG algebra, we proceed as in the twisted Sugawara construction and define\footnote{For generic $\vk$, we would have $L(\vec{\phi})=\frac{1}{|k|^2}\beta(\vec{\phi}) \vk\cdot\vec{\partial} (\vk\cdot\vec{J}(\vec{\phi})) +\dfrac{1}{2}g^{ij} J_i(\vec{\phi})J_j(\vec{\phi})$
with $ \vk\cdot\vec{\partial}\beta(\vec{\phi}) =0$.}
\be
\label{L-J-Sugawara}
L(\phii,\Phi)
\equiv
\beta(\Phi)\, k^i J'_i(\phii, \Phi)+\dfrac{1}{2}g^{ij} J_i(\phii, \Phi)J_j(\phii, \Phi)
\ee
where $g^{ij}$ is the inverse of the metric $g_{ij}$, $\beta (\Phi)$ is an arbitrary function of $\Phi$ and as before the prime denotes partial derivative with respect to $\phii$. One can readily see that the the modes $L_n(\Phi)$ of this quantity close according to the NHEG algebra \eqref{Vir-bundle-algebra} with a $\Phi$-dependent central charge
\be
\label{ciphi}
c(\Phi)
=
12\,\beta^2(\Phi)\,k^i g_{ij} k^jZ.
\ee
Note that this is non-zero only if the twisting term $\beta kJ'$ in (\ref{L-J-Sugawara}) does not vanish. As for the brackets of NHEG generators with currents, they take the form
\be
\label{L-J-algebra}
i[L_\vm, J_{i,\vn}]=-(\vk\cdot\vn)J_{i,\vm+\vn}+i\beta(\vk\cdot\vn)^2k_i \, \de_{\vm+\vn,0}\,Z.
\ee
The central term on the right-hand side implies that the currents $J_i$ are not primary fields; this is of course a standard feature of the twisted Sugawara construction (see \eg \cite{Afshar-et-al, AGSY}), where the twist term proportional to $J'$ in (\ref{L-J-Sugawara}) gives rise both to the classical central extension (\ref{ciphi}) and to the anomaly in \eqref{L-J-algebra}. In the next section we will use this free-field point of view to build NHEG representations.

\newpage
\section{Quantization of NHEG orbits}
\label{Sec6}

So far we have analysed NHEG symmetry from a classical perspective involving symplectic manifolds and Poisson brackets. We now study some aspects of the quantization of these orbits, whereby Poisson brackets are replaced by commutators according to the canonical prescription $i \{\cdot, \cdot\}\to [\cdot,\cdot]$. This results in a Hilbert space acted upon by operators $L_{n}(\Phi)$ that satisfy the commutator algebra
\be\label{NHEG-algebra-quantized}
\big[L_m(\Phi),L_n(\Phi')\big]=
\left((m-n)L_{m+n}(\Phi) +\frac{c(\Phi)}{12}(m^3-m)\delta_{m+n,0}\right)\delta^{n}(\Phi-\Phi').
\ee
Here we have shifted the $L_0(\Phi)$ of (\ref{Vir-bundle-algebra}) by $-c(\Phi)/24$, and we assume that the central charge is some given, strictly positive function $c(\Phi)$. The operators $L_n(\Phi)$ satisfy the Hermiticity conditions $L_m(\Phi)^{\dagger}=L_{-m}(\Phi)$. Similarly, the quantization of the current algebra (\ref{J-Phi-algebra}) gives commutators
\be
[J_{i,m}(\Phi),J_{j,n}(\Phi')]= m\, g_{ij}\,\de_{m+n,0}\ \delta^{n}(\Phi-\Phi'),
\ee
where $J_{i,m}(\Phi)^\dagger=J_{i,-m}(\Phi)$. In the remainder of this section we describe irreducible unitary representations where these commutator algebras are realized.

\subsection{NHEG unitarity}
\label{NEGU}

Since NHEG orbits are bundles of Virasoro orbits over $T^n$, their quantization is in principle straightforward if one assumes that the quantization of Virasoro orbits goes through.\footnote{The quantization of Virasoro orbits is still very much an area of active research; see \eg \cite{Salmasian:2014wwa}. Our standard of rigour is by no means that of pure mathematics, so we shall bluntly assume that Virasoro quantization does work.} Indeed, if the Hilbert space obtained by quantizing a Virasoro orbit is the space of a unitary highest-weight representation of the Virasoro algebra, then the Hilbert space obtained by quantizing a NHEG orbit is a continuous tensor product of Virasoro representations --- one at each point $\Phi$ of $T^{n}$. In particular, the highest-weight state $|h\rangle$ of a NHEG representation is specified by a strictly positive real function $h(\Phi)$ on $T^{n}$. It is such that
\be
L_0(\Phi)|h\rangle=h(\Phi)|h\rangle,
\qquad
L_m(\Phi)|h\rangle=0
\qquad
\forall\Phi\in T^{n},\;\forall m>0,
\ee
and we assume it to be normalized: $\langle h|h\rangle=1$. The NHEG vacuum state $|0\rangle$ is specified by $h(\Phi)=0$ for all $\Phi\in T^n$, and it is annihilated by all $L_m(\Phi)$'s with $m\geq-1$.

Any descendant of the highest-weight state is specified by $N$ insertion points $\Phi_1,...,\Phi_N$ on the torus, and takes the form
\be
\label{Deka}
L_{-n^1_1}(\Phi_1)L_{-n^1_2}(\Phi_1)...L_{-n^1_{k_1}}(\Phi_1)
...
L_{-n^N_1}(\Phi_N)...L_{-n^N_{k_N}}(\Phi_N)|h\rangle
\ee
where $1\leq n^i_1\leq...\leq n^i_{k_i}$ for all $i=1,...,N$. Strictly speaking, the norm of any such descendant is infinite due to the delta function on the right-hand side of (\ref{NHEG-algebra-quantized}). For instance,
\be
\langle h|
L_m(\Phi)L_{-m}(\Phi')
|h\rangle
=
\Big[2mh(\Phi)+\frac{c(\Phi)}{12}(m^3-m)\Big]\delta^n(\Phi-\Phi'),
\ee
so one can think of these descendants as analogues of states with definite position in non-relativistic quantum mechanics. The true elements of the Hilbert space are actually smeared linear combinations of descendants, such as
\be
\int_{T^n}\!\!d\Phi\,\Psi(\Phi)L_{-m}(\Phi)|h\rangle
\label{WaFa}
\ee
where $\Psi$ is a square-integrable wavefunction on $T^n$. More generally, the number of wavefunctions needed to smear a descendant (\ref{Deka}) is the number $\sum_{i=1}^Nk_i$ of $L_m(\Phi)$ operators appearing in its expression. In particular, smeared states are generally highly non-local on $T^n$.

As in the Verma modules of the Virasoro algebra, the Hilbert space $\mathscr{H}$ of the representation is spanned by all linear combinations of the highest-weight state and its descendants; here we are including smearing such as (\ref{WaFa}) as a valid way to take linear combinations. Thus, abstractly, the Hilbert space is an infinite tensor product of Virasoro Verma modules $\mathbb{V}_{c(\Phi),h(\Phi)}$, one at each point of the $n$-torus:
\be
\mathscr{H}=\bigotimes_{\Phi\in T^n}\mathbb{V}_{c(\Phi),h(\Phi)}.
\label{tensor}
\ee
This product is the quantization of eq.\ (\ref{Obobo}); it is separable thanks to the fact that $\Phi$ is a coordinate on a torus rather than a non-compact manifold. In principle, any irreducible unitary representation of the NHEG algebra takes this form, and similar conclusions apply more generally to any quantum theory with NHEG symmetry: any such theory is a bundle (\ie a tensor product) of chiral two-dimensional CFTs over an $n$-torus. One might say that $T^n$ is a ``conformal manifold'', or moduli space, supporting a family of CFTs; the novelty here is that (i) this manifold has a space-time interpretation since it is embedded in the NHEG (\ref{NHEG-metric-generic}), and (ii) the symmetry algebra has this moduli space built in, resulting in wavefunctions such as (\ref{WaFa}) that live on the conformal manifold.

\paragraph{Hilbert space from NHEG-Kac-Moody.} Unitary representations of the NHEG algebra can also be built thanks to the currents $J_i$ of section \ref{secSug}. As in the Virasoro case \cite{AGSY}, these currents can be seen as creation/annihilation operators generating the space of the representation when they act on a suitable vacuum state. To define the latter, we start by noting that
\be
[J_{i,0},L_\vm]=0=[J_{i,0},J_{j,\vm}],\qquad \forall \vm\in\mathbb{Z}^{n+1},\,\forall i,j=1,...,n+1,
\ee
implying that the $n$ zero-modes $\bJ_{i,0}$ commute with all other generators; in fact, they span the center of the universal enveloping algebra of NHEG-Kac-Moody. Accordingly, we label the vacuum state $|J_1,...,J_{n+1}\rangle\equiv|J_i\rangle$ by its (local) eigenvalues $J_i(\Phi)$ under central operators:
\be
J_{i,m}(\Phi) |J_i\rangle=0
\quad\text{and}\quad
J_{i,0}(\Phi)|J_i\rangle= J_i (\Phi)|J_i\rangle
\qquad\forall m>0,\ \forall\Phi\in T^{n}.
\ee
From the point of view of the NHEG algebra obtained by normal-ordering \eqref{L-J-Sugawara} and shifting it by $c(\Phi)/24$, these vacua are primary states of $\Phi$-dependent weight $h(\Phi)=\frac{c(\Phi)}{24}+\frac12 g^{ij} J_i(\Phi)J_j(\Phi)$:
\be\label{L-vacuum}
L_{m}(\Phi) |J_k\rangle=0 \ \ \forall m>0,
\qquad
L_0(\Phi)|J_k\rangle=h(\Phi) |J_k\rangle.
\ee
Each ``vacuum'' of NHEG-Kac-Moody is thus specified by $n+1$ functions $J_i(\Phi)$ on $T^n$. Once such a vacuum has been chosen, its descendants are built by acting with creation operators $J_{i,-n}(\Phi),\ n>0$, at generally different insertion points on $T^n$. This is directly analogous to the construction (\ref{Deka}) in terms of NHEG generators, and one can indeed show (along the same lines as for the Virasoro case \cite{AGSY}) that the Hilbert space spanned by descendants of $|h\rangle$ is isomorphic, as a NHEG-module, to the one spanned by descendants of $|J_i\rangle$ provided $h(\Phi)\geq0$ at any point $\Phi$ of $T^{n}$.

\subsection{NHEG characters}

To conclude this section, let us evaluate the characters of the representations above. First note that the Cartan subalgebra of NHEG is infinite-dimensional: as is apparent in (\ref{NHEG-algebra-quantized}), it is spanned by all generators $\bL_0(\Phi)$ with $\Phi\in T^{n}$. This implies that the chemical potential itself is generally a function of $\Phi$, $\tau(\Phi)$; as in standard CFT we assume it has positive imaginary part everywhere. The associated character is
\be
\chi_{h,c}(\tau)
=
\text{Tr}_{\mathscr{H}}\Big(
e^{\int_{T^{n}}\!\!d\Phi\, 2\pi i\tau(\Phi)\,L_0(\Phi)}
\Big)
\stackrel{\text{(\ref{tensor})}}{=}
\prod_{\Phi\in T^{n}}\text{Tr}_{\mathbb{V}_{c(\Phi),h(\Phi)}}\Big(e^{2\pi i\tau(\Phi)\,L_0(\Phi)}\Big),
\label{chachacha}
\ee
where $d\Phi\equiv d\phi^1...d\phi^n/(2\pi)^n$ as before. For definiteness we assume that $c(\Phi)>1$ and $h(\Phi)>0$ everywhere on $T^n$. Then there are no null states in $\mathbb{V}_{c(\Phi),h(\Phi)}$ and the counting of $L_0$ eigenstates at level $N$ reduces to the counting of partitions $p(N)$ of the integer $N$.  Using this in (\ref{chachacha}), we conclude that
\be
\chi_{h,c}(\tau)
=
\prod_{\Phi\in T^{n}}
e^{2\pi i\tau(\Phi)\,h(\Phi)}\prod_{m=1}^{\infty}\frac{1}{1-e^{2\pi im\tau(\Phi)}}
=
\exp\bigg[\int_{T^{n}}\!\!\!\!d\Phi\bigg(2\pi i\tau(\Phi)\,h(\Phi)-\sum_{m=1}^{\infty}\log\big(1-e^{2\pi im\tau(\Phi)}\big)\bigg)\bigg].
\label{samba}
\ee
A similar counting works for the vacuum representation where $h(\Phi)=0$ for all $\Phi$, but then the product and sum over $m$ start at $m=2$. In the special case where $\tau(\Phi)$ is constant, the torus integral in (\ref{samba}) is finite ($\int d\Phi=1$) and the character reduces to
\be
\chi_{h,c}(\tau)
=
q^{\,\int d\Phi\,h(\Phi)}
\prod_{m=1}^{\infty}\frac{1}{1-q^{m}},
\qquad
q\equiv e^{2\pi i\tau}.
\label{sambapati}
\ee
This is just the character of a standard unitary Virasoro representation, free of null states, whose effective highest weight $h_{\text{eff}}=\int d\Phi\,h(\Phi)$ is the average of the $h(\Phi)$'s. (Again, when $h=0$ the product would start at $m=2$.) In principle one can perform the same computation when $0<c(\Phi)<1$, in which case the local Virasoro representations contain null states and the weights $h(\Phi)$ are constrained by the Kac determinant. Then both $c$ and $h$ are forced to take discrete values, implying that they are both constant over $T^{n}$ if one assumes continuity. The corresponding NHEG character thus coincides with the character of a reducible Verma module.

Recalling the relation between characters and gravitational partition functions \cite{Giombi:2008vd}, it would be interesting to see if the NHEG characters (\ref{samba})-(\ref{sambapati}) have anything to do with (one-loop) partition functions of the gravitational field in the near-horizon region of extremal black holes. We will not attempt to address this intriguing issue here.

\section{Discussion and outlook}
\label{secCon}

In this work we have built the NHEG group, starting from the NHEG algebra presented in \cite{CHSS-1,CHSS-2}. As we have seen, the NHEG group consists of diffeomorphisms of an $(n+1)$-torus that preserve the direction of an anisotropy vector $\vk$. When the latter is proportional to an element of the dual lattice, the NHEG group effectively becomes a bundle of Virasoro groups over an $n$-torus, generally with a position-dependent central charge. This simple observation allowed us to derive, for free, the classification of NHEG coadjoint orbits: each such orbit is a bundle of Virasoro orbits over $T^n$ (see fig.\ \ref{FIG}); the Virasoro orbits at different points generally differ, but they are constrained by the requirement that they form a smooth bundle, leading for instance to a statement of unicity of the vacuum NHEG orbit. In addition we derived the corresponding irreducible unitary representations, which are simply tensor products of Virasoro modules, and we computed their characters. In the remainder of this section we briefly address some open issues, extensions and plausible applications of our analysis.

\paragraph{Orbits versus Metrics.} As reviewed in section \ref{secNHEG}, the NHEG group (with constant central charge) represents the symplectic symmetries of the phase space of metrics \eqref{g-F}. This space is a homogeneous manifold with a NHEG-invariant symplectic form, so it is one of the coadjoint orbits described in section \ref{secOrb}; in fact it is the (hyperbolic) orbit of the constant coadjoint vector $L_{\Phi}=c/12$ with $c=6S/\pi$, provided one identifies (\ref{Scharz transform}) with $\widehat{\text{Ad}}{}^*_{\cF^{-1}}\bL$ up to an overall normalization. This parallels similar identifications between AdS$_3$ metrics and Virasoro coadjoint orbits \cite{CMSS, Garbarz:2014kaa,Barnich:2014zoa}, or their flat space/BMS$_3$ analogues \cite{Barnich:2015uva}. In the same vein, one may ask whether more interesting profiles of NHEG coadjoint vectors, \eg those described in  (\ref{exeppe}), correspond to well-defined space-time metrics solving the vacuum Einstein equations.

Unfortunately, a geometric subtlety makes this identification between NHEG coadjoint orbits and orbits of NHEG metrics somewhat difficult. Indeed, our construction of the NHEG group (and of its orbits) used the fact that the anisotropy vector $\vk$ can be transformed in such a way that $\vk=(0,...,0,1)$; but in order for this to work in the case of NHEG metrics, one would have to perform a similar transformation at all values of the azimuthal angle $\theta$ (recall the coordinates used when writing (\ref{NHEG-metric-generic})). In general, such a $\theta$-dependent transformation is singular, thus preventing a direct identification between NHEG coadjoint vectors $L_{\Phi}(\phii)$ and NHEG metrics. A notable exception to this general expectation is provided by coadjoint orbits whose representative element is a constant, \ie those with  $L_{\Phi}(\phii)=const$. Intuitively, this is consistent with the na\"ive expectation that labels of NHEG orbits should be related to the conserved charges associated with exact symmetries/Killing vectors of the background geometry (see \cite{CHSS-2,CMSS,HS}). Whether other NHEG orbits correspond to smooth space-time metrics is a puzzling question, and we hope to return to it in the future.

\paragraph{More on NHEG-Kac-Moody.} In section \ref{secSug} we built the NHEG algebra in terms of u(1) currents thanks to a twisted Sugawara construction, which we then applied to unitary NHEG representations along the same lines as in the Virasoro case \cite{AGSY}. In that context, a natural question is whether the currents have a geometric realization in terms of space-time diffeomorphisms, as is the case for instance in three dimensions \cite{Afshar-et-al}. This project has already been carried out for extreme Kerr black holes \cite{Extreme-Kerr-Fluff}, but it should be possible to extend it to higher dimensions.

A preliminary analysis extending the method of \cite{Extreme-Kerr-Fluff} has uncovered a closely related algebra whose generators are u(1) currents that depend on individual $\phi^i$'s, from which one can obtain $n+1$ copies of Virasoro algebras \cite{NHEG-Fluff}; it would be interesting to explore this direction further.

\paragraph{NHEG Field Theories.} As mentioned in section \ref{NEGU}, NHEG-invariant field theories are bundles of CFTs over a $T^n$ moduli space. The point of view adopted in this paper, and the motivation for our investigation, is that such field theories provide putative holographic duals for extremal black holes. It would be interesting to see what constraints are put on such theories by the requirement that they describe gravity. An obvious constraint comes from the Poisson bracket algebra (\ref{sufacha}), which says that the position-dependent central NHEG charge must in fact be a constant (at least up to quantum corrections). Optimistically, a better understanding of NHEG field theories might eventually lead to an identification of extremal black hole microstates that does not rely on string theory \cite{Strominger:1996sh}.

An illustrative application of the considerations of this paper to the identification of black hole microstates is provided by the recent ``horizon fluff'' proposal \cite{AGSY}, according to which these microstates can be read off from a subtle relationship between {\it near-horizon} and {\it asymptotic} symmetry algebras. This proposal has been successfully worked out for generic BTZ black holes\cite{AGSY} and extremal Kerr black holes \cite{Extreme-Kerr-Fluff}, and is expected to work for NHEGs too \cite{NHEG-Fluff}. From this perspective, the orbit analysis presented in this work should be an important technical tool for the exploration of black hole entropy.

\section*{Acknowledgements}

We are especially grateful to Ali Seraj for his contributions in early stages of this project, and we also thank Geoffrey Comp\`ere for comments and discussions. The work of B.O.\ is supported by the Swiss National Science Foundation, and partly by the NCCR SwissMAP. M.M.Sh-J.\ would like to thank Abdus Salam ICTP where part of this work was carried out and he acknowledges the ICTP Simons associates program and the ICTP NT-04 network scheme. He also acknowledges support from the Iranian NSF junior research chair in black hole physics.

\end{document}